# Quantifying Collective Memories

Cristian Candia[1,2,*] (0000-0002-9654-543X)

[1]Data Science Institute, Facultad de Ingeniería, Universidad del Desarrollo, Las Condes, 7610658, Chile

[2]Northwestern Institute on Complex Systems (NICO), Northwestern University, Evanston, IL 60208

* Corresponding Author: crcandiav@gmail.com

**Abstract (150 words)**

Collective memory is a common representation of the past shared by a group of people that modulates its identity. Recent literature on computational social science quantifies collective memories using expressions of those memories operationalized as the amount of collective attention focused on specific cultural icons, either artifacts or people. We model the temporal dynamics of collective memory using a two-step decay process, characterized by a short-lived and intense initial stage followed by a long-lived and milder decline of collective attention. Different collective memory mechanisms sustain the two-step process. The first is communicative memory, which corresponds to all memories supported by socializing acts. The second is cultural memory, which corresponds to all memories sustained by accessing records. Thus, this model predicts a transition time when cultural memory overcomes communicative memory. It has practical consequences related to shrinking or extending the time a specific topic is part of a community's conversation, i.e., communicative memory, empowering policymakers with valuable time to generate solutions and compromises for contingent and public problems.

## Introduction

Collective memory is the common representation of the past shared by a group of people that shapes their cultural identity (Assmann, 1995; Hirst & Manier, 2008). Collective memory can be divided into two types: Communicative memory, meaning all the memories sustained by acts of socialization, and cultural memory, meaning all the memories maintained by access to records. In other words, collective memories are all the information, memories, and knowledge sustained by communities (as small as a family or as large as all the speakers of a language) that at the same time shape communities' identities (Hirst & Manier, 2008). Communities keep their collective memories alive by two processes: communicative memory and cultural memory. Researchers studying collective memory have contributed a variety of mechanisms, definitions, and techniques to characterize different forms of collective memory and the determinants for their preservation (Hirst et al., 2018).

Collective memory (Assmann, 1995) allows us to build the past in a two-fold way. On the one hand, as a social construction, and on the other, through vestiges, personal memories, relics, or archives (Nora, 1989). Thus, collective memory contributes to finding an identity characteristic of the human being. This identity allows us to orient ourselves in time and space, thinking about time horizons much further away than our birth or death (Assmann, 2011). Assmann states that there are three levels in which the connection between memory, time, and identity operates: (i) Internal or individual level, where only the individual's memory operates. (ii) Social level, where memory is about communication and social interactions; memory allows us to live in a community, and the community allows us to build a memory. (iii) The cultural level, initially proposed by the art historian Aby Warburg, dealt with images (culture in the form of objects: music, art, and scientific pieces) as memory carriers.

On a different strand of literature, Hirst et al. (2018) propose an individual and subjective representation of memories, which tells us how collective memories are subjectively represented. Individuals remember collective memories in their minds from perceptions and thoughts and not from the physical (objects, books, or monuments, among others). This point is also approached in two different ways, with bottom-up or top-down mechanisms. Researchers who work with a top-down approach look at the expressions of collective memory and then investigate what cognitive processes form it. They are interested in understanding how specific (generally national) collective memories are formed. While researchers who work with a bottom-up approach focus on the cognitive processes that could lead to the formation and maintenance of specific collective memories, interested in how these memories come to be shared, without reference to existing collective memories.

Top-down mechanisms depend on the existing context of the community (Hirst et al., 2018) and how it contributes to the retention and formation of collective memories such as narrative templates (Hammack, 2011; Wertsch, 2002), familiarity (Roediger & DeSoto, 2016; Rubin, 1995), and cultural attractors (Buskell, 2017; Richerson & Boyd, 2005; Sperber & Hirschfeld, 2004). For instance, familiarity increases the memorability of events, even causing false memories, like that of people identifying Alexander Hamilton as a U.S. president (Roediger & DeSoto, 2016). Narrative templates, which are schemata that people use to describe multiple historical events, can also shape memories, such as the memory of Russian exceptionalism that emerges from the narrative template of invasion, near defeat, and heroic triumph (Wertsch, 2002). Cultural attractors, such as repetitive children's songs or count-out rhymes, can increase the preservation of memories across generations (Rubin, 1995). Bottom-up mechanisms depend on micro-level psychological processes that drive social outcomes (Hirst et al., 2018), such as retrieval-induced forgetting (Cuc et al., 2007; Garcia-Bajos et al., 2009; Storm et al., 2012), and social affinities, which increase the mnemonic power of conversations (Coman et al., 2014; Coman & Hirst, 2015; Echterhoff et al., 2009; Stone et al., 2010). For instance, people share realities with those they perceive as belonging to their group (Coman & Hirst, 2015).

In summary, memory cannot be understood without considering the group to which people belong and the social contexts within which we evolve as individuals (Halbwachs, 1950).

Also, collective memory is dynamic. It constantly undergoes transformations and transitions over time. It remains in constant evolution due to generational changes, and because memories are increasingly blurred in people's minds, so it is distorted and manipulated with the passing of the years. One of these transformations, and perhaps the most obvious, is the transition from communicative memory that allows us to socialize and transmit knowledge to cultural memory that functions as a repository of socially constructed stories (Assmann, 2011; Confino, 1997; Nora, 1989).

## Communicative Memory

Communicative memory is defined by Assmann (1995, 2008) as a type of collective memory based on daily communication, characterized by high informality, thematic instability, and disorganization. This type of memory can be observed in any day-to-day conversation, changing from topic to topic without the need to address these topics in depth or in a knowledgeable way. For this reason, the time horizon of communicative memory is limited. However, through these communicative acts, it is possible to create a socially mediated memory related to a group, which, in turn, has a way of seeing its past similarly.

Let us recall the bottom-up approach proposed by Hirst et al (2018), which assumes that the processes of individual memory create a social result. Here, collective memory is an epidemiological process, where communication is the virus that spreads and infects the rest of the information and memories. These are inevitably selective and low fidelity memories, which can cause a variation in the memory of both the speaker and the listeners. This approach has to do with how people relate their own life stories to relevant historical events; if someone       personally experienced this event, it means more to them than to others. If, therefore, this fact is transmitted to the next generation, it also acquires a higher degree of significance than transmitting the message to a subsequent generation. So, collective memory also involves events that occur during people's lives and are part of their history. The theory of temporal construction establishes that the more psychologically distant an object is from an individual, the more abstract its representation is and vice versa. It is natural to say that experiences are usually remembered with personal and specific acts. In contrast, not experienced facts are typically described in an impersonal and much more expansive way. These memories are transmitted through conversations, so their impact on the listener depends exclusively on who tells the story and their representation of it.

Communicative memory does not have physical aspects capable of recalling but is found in everyday interactions and communications. For this reason, it has a limited duration, and this duration depends on social ties and time frames (Assmann, 2011). We will understand time frames, as a framework from the recent past, which refers to social or communicative memory, it is gradually forgotten over time and does not last more than three generations. On the other hand, conversational influences are represented by three major stages: first to reinforce knowledge; second, to replace knowledge with new information; and third, to induce forgetting (Hirst et al., 2018). Thus, people are open to talking to people who think similarly or belong to the same social group; this reinforces their knowledge and gives space to "forget" pre-existing beliefs.

Researchers of the bottom-up approach (Hirst et al., 2018) measure memories in large groups, for example, how conversational influences spread, causing mnemonic convergences. They also explored how network typologies affect this convergence, concluding that for non-clustered networks, the mnemonic convergence was higher than for clustered networks (Coman et al., 2016). In other words, the low fidelity of human memory, which can be characterized as a weakness, is likely to be a great strength since, as it spreads through conversation and mnemonic convergence, it generates a "universal truth," generating cultural memory.

## Cultural memory

Although the concept of cultural memory has been explored since the 1930s, it was only in the 1980s that the term was coined, and the connection between time, identity, and memory became present at its three levels: personal, social, and collective. Cultural memory is a form of collective memory that has been defined as the memory that is generated by the connections between human beings building a joint memory (Halbwash, 1992), and a memory where the iconic is an essential part of memory (Assmann, 2011).

Cultural memory is characterized by being distant from the daily; that is, its starting point or horizon does not change over time. It is maintained and it is permanent. Generally, these starting points are fateful situations of the past that are kept "alive" as memory figures; these memory figures correspond to textbooks, rituals, monuments, among others (Assmann, 1995). Cultural memory is externalized, it is objective, and it is stored in symbolic forms that are stable and transcendent; that is, they are capable of being transferred from generation to generation; this would be the case of books, songs, patents, or movies (Assmann, 2011). For them to be truly considered memory, they must be able to circulate and take root in a society. The objective of all these pieces is not that they possess memory by themselves but that they can invoke memories in those who see them. So, while this memory is stored in symbolic, objective forms, the memories they invoke are entirely subjective and depend on who is looking at the symbols.

Some relevant characteristics of the cultural memory (Assmann, 1995) are related to: i) the creation of identity, where the group identifies itself and determines if it belongs or not, ii) the ability to reconstruct, although figures of memory perpetuate an immovable horizon, the interpretation that each generation makes of them may be different in each era, iii) stable formation through means such as writing, images, and rites, iv) organization, which on the one hand forces to communicate the facts as time passes and the high specialization of memory bearers, v) the obligation that refers to the creation of a system of values and differentiation and finally, vi ) reflection through reflective-practice (rituals), self-reflection (recourses to ourself to explain ourself) and reflection to our image (concern for our social system).

## Collective Memory and Attention

Halbwachs in 1925 stated, *"the idea of an individual memory, absolutely separate from social memory, is an abstraction almost devoid of meaning"* (Connerton, 1989; Halbwachs, 1992). However, the assertion that individual memory cannot be conceptualized and studied without considering its social context does not imply the reverse (Kansteiner, 2002). This means that collective memory can only be conceived and accessed through its expression in individuals. Hence, it is necessary to differentiate between autobiographical memory and collective memory. For lack of such differentiation, many investigations regarding collective memories commit an attractive yet potentially serious methodological error: they perceive and conceptualize collective memory exclusively in terms of the psychological and emotional dynamics of individual remembering (Kansteiner, 2002). Classic literature about collective memory has been focused on the memories of individuals and then aggregated the data (Hill et al., 2013; Roediger et al., 2009; Roediger & DeSoto, 2014; Rubin, 1998, 2014; Rubin & Wenzel, 1996; Wixted & Ebbesen, 1997; Zaromb et al., 2014). On this line, Jeffrey Olick, cited by Kansteiner (2002), offered a distinction between "collected" and "collective" memory. A collected memory is the traditional aggregation of individual data. Still, collective memory does not have the same dynamics, so, Kansteiner states that it is necessary to find another suitable method in order to study collective memory.

Here, we use a computational social science approach, which can capture the underlying context of the studied communities, given the nature of the used data. The data carries the collective behavior of people in the respective system because it is an expression of their individual decision-making process that translates in focusing collective attention into cultural pieces and icons, for instance, citing a paper or patent, listening to a song, or watching a movie.

Collective attention is pivotal for decision-making and for the networked spread of information and ideas. Studies on collective attention have been mainly conducted at the individual level and small group level (Wu & Huberman, 2007) by psychologists (Kahneman, 1973; Pashler, 1998), economists (Camerer, 2003), and marketing researchers (Dukas, 2004; Pieters et al., 1999; Reis, 2006). An essential part of the progress in collective attention research is in the theoretical literature of attention economics and in small laboratories (Falkinger, 2007; Simon, 1971). Therefore, collective attention research needs more quantitative models to connect it empirically to large-scale data.

Nowadays, practical applications of collective attention research—ranging from the prediction of life cycles of a movie (Candia et al., 2019; Hidalgo et al., 2006) to the resilience of a country in a traumatic experience (Ferron & Massa, 2014; Klep, 2012; Truc, 2011) (e.g., September 11th)—make collective attention of particular interest for researchers from a variety of research fields. Thus, recent literature on computational social science has shown several advances on quantitative models in different objects of study (Candia et al., 2019; Candia & Uzzi, 2021; Coman, 2019; Cunico et al., 2021; García-Gavilanes et al., 2016, 2017; Higham et al., 2017b, 2017a; Jara-Figueroa et al., 2019; Kanhabua et al., 2014; Lorenz-Spreen et al., 2019; Mukherjee, Romero, et al., 2017; Mukherjee, Uzzi, et al., 2017; Ronen et al., 2014; Sinatra et al., 2016; Skiena & Ward, 2014; Uzzi et al., 2013; Wang et al., 2013; Yu et al., 2016; Yucesoy & Barabási, 2016).

Concretely, the empirical literature on collective attention models people's engagement (adoption, consumption, or diffusion) to cultural content as a combination of two processes. First, is the growth in the number of people attending to a given cultural piece (preferential attachment or cumulative advantage mechanisms). Second, as time goes on, the habituation or competition with other cultural pieces provoke       that old cultural pieces will be less attractive to people (temporal decay or obsolescence mechanism) (Candia et al., 2019; Csárdi et al., 2007; Dorogovtsev & Mendes, 2000; Golosovsky & Solomon, 2012; Higham et al., 2017a, 2017b; Valverde et al., 2007; Wang et al., 2013).

Preferential attachment (Albert & Barabasi, 2002; Barabási & Albert, 1999) or cumulative advantage (Merton, 1988; Price, 1976; Yule, 1925) refers to the process in which attention begets attention. Think of two computational social scientists' papers, one with ten citations and another one with 100 at a given time, t. The probability that the second paper receives a new citation at the time t+1 is more significant than the first one, simply because more people already know about the highly cited article; thus, generating growth of current attention.

Yet, this cumulative advantaged growth competes with temporal decay. Think about the current collective attention of movies. The top-10 most-watched movies in 2017 cannot compete with the top-10 most-watched movies in 2021. Even though they achieved the same accomplishments (get into the top-ranked movies in their releasing years), the current collective attention (e.g., the number of people consuming those movies) is higher for 2021's movies than past movies. Given society's limited and finite collective attention capacity, the novelty of movies decreases as time passes because of habituation or competition with other new films. Thus, the collective attention that movies (and cultural pieces and icons in general) receive decreases over time (Falkinger, 2007; Wu & Huberman, 2007).

The temporal decay has been modeled using different functional forms, from simple exponentials to log-normal distributions and power-laws. Recent literature shows that the collective attention temporal decay of cultural pieces (songs, movies, patents, and scientific articles) and cultural icons (i.e., globally famous people) decay in a two-step process (Candia et al., 2019; Candia & Uzzi, 2021; Higham et al., 2017b, 2017a; West et al., 2021). For a group of cultural pieces created in a given year, the first stage is attentionally intensive and short-lived, i.e., people focus a high portion of their collective attention on them but for a short time. The second stage is attentionally mild and long-lived, i.e., people focus a low amount of their collective attention on them for a long time.

To explain the two-step temporal decay of collective attention, we rely on collective memory mechanisms. Figure 1 shows the temporal decay process of collective memory and attention. The portion of collective memory, focused on a particular set of cultural pieces, decreases over time following two paths: i) Some of the memories sustained by communicative acts (communicative memories) decay directly to oblivion, quantified as the parameter p. ii) The rest of the communicative memories become memories sustained by accessing records (cultural memories), quantified as the parameter r. Finally, cultural memories also decay to oblivion, quantified as the parameter q, but slower than communicative memories that decay directly to oblivion. The red curve, which follows Candia et al., 2019 model, reproduces the two-step temporal decay process where the first part is driven by communicative memory (purple line) and the second part is mainly driven by cultural memory (blue curve).

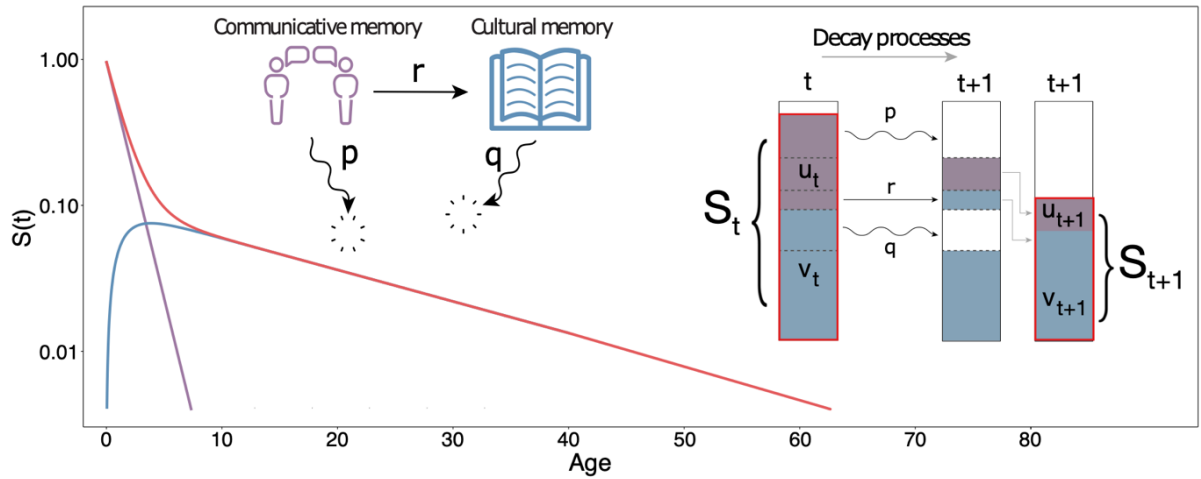

*Figure 1: The y axis represents the normalized current level of attention received by a group of comparable cultural pieces. The x-axis represents the age of the cultural pieces. The red curve shows the biexponential function predicted by our model on a log-linear scale. The blue and purple curves show the two exponentials of communicative and cultural memory. The inset illustrates the basic mechanics of the model. At any time, t, the total memory is the sum of communicative memory u and cultural memory v. Both communicative and cultural memory decay with their own respective decay rates p + r and q, and cultural memory grows with r. Note. Figure reprinted from "The universal decay of collective memory and attention", by Candia et al., 2019.*

## The Link Between Collective Memory and Collective Attention

To exemplify how collective memory and collective attention are associated, let us consider the following conceptual example. In a college library, we can identify three distinct areas by the number of students consulting books (i.e., levels of collective attention): i) The front part shelters the most recent textbooks and other recurrent class material. On-demand, new versions of textbooks come out periodically, including reviews and related pieces of content. ii) The back part contains older books, or probably recently replaced versions of old textbooks, thus replacing them in the front part. Smaller groups of students consult those books. iii) The basement: The books that live here are in the catalog, but they are hardly being consulted because, for instance, a new paradigm came out or just because people do not have an interest in them anymore. However, they can become relevant if something changes in the system.

The analogy of the college library enlightens the difference and relationship between actively remembered knowledge or, in Assmann's terms (2008), the canon,"actively circulated memory that keeps the past present" (p.98), and the archive, "passively stored memory that preserves the past" (p.98). Memory studies' scholars have extensively argued that culture is inextricably linked to memory (Assmann, 1995; Halbwachs, 1992, 1997) and its parallel, forgetting (Assmann, 2008). Furthermore, Assmann (1995) defines two modes of cultural memory "... the mode of potentiality of the archive whose accumulated texts, images, and rules of conduct act as a total horizon, and ... the mode of actuality, whereby each contemporary context puts the objectivized meaning into its own perspective, giving it its own relevance." (p.130). The mode of potentiality corresponds to the existence of records (e.g., old books in a library basement that are hardly consulted), and the mode of actuality corresponds to the pieces whose records receive attention when they become relevant to the community. Any cultural piece in a mode of potentiality can transit to a mode of actuality if the community decides that the piece is important.

We can argue that the first level of attention, described in the library example, is mainly shaped by communicative acts, e.g., classes, that focus the attention of specific communities on specific pieces of information (Candia et al., 2019). Cultural pieces that receive the most collective attention live in a mode of actuality (Assmann, 1995), form part of the canon (Assmann, 2008), and their current interpretation is part of the community's identity. The second level of attention is mostly characterized by accessing records, e.g., literature searching,      and the access can be related to both bottom-up and top-down mechanisms. We can say these cultural products live in a midpoint between the mode of potentiality and the mode of actuality. The third level of attention is close to zero. We reasonably argue that those books, in Assmann's words, live in the potentiality mode (1995). These books are stored in an archive and are not part of the everyday conversation of the community, but they can be retrieved if the community decides to remember those pieces. Hence, remembering and forgetting are dynamic. As Assmann (2008) argues, "[e]lements of the canon can ... recede into the archive, while elements of the archive may be recovered and reclaimed for the canon." (p. 104).

Thus, the computational social science approach works close to Assmann's definition of collective memory (1995, 2011), which focuses on the cultural products that communities or groups of people remember and focus their collective attention on (Fig. 1). Assmann focused on long-lived inter-generational memories, although he distinguishes between modes of potentiality and actuality.

## Empirical Strategy

Measuring the expressions of collective memories through how people focus their collective attention on certain cultural pieces is pivotal for quantifying collective memories. Following the empirical literature of collective attention, the driving forces of collective attention are the growth in the number of people that attend to a given cultural piece (preferential attachment or cumulative advantage mechanisms) and the habituation or competition with other cultural pieces that make the same cultural piece less likely to be attractive as time goes on (temporal decay or obsolescence mechanism) (Candia et al., 2019; Candia & Uzzi, 2021).

First, we need to start tracking collective attention over time for a given community. Without loss of generality, let us consider an example using citation data to scientific articles in the Physical Review B (PRB) journal published by the American Physical Society (APA), which defines the community. Figure 2A shows two different approaches to tracking collective attention. The retrospective approach (dashed arrows, Figure 2A) considers papers cited by a publication during a particular year and then analyzes the distribution of their ages retrospectively (Bouabid, 2011; Burrell, 2002; Glänzel, 2004). The retrospective approach looks at the past, i.e., it focuses on tracking the referenced papers of a given cohort. The prospective method (solid arrows, Figure 2A) analyzes the distribution of citations gained over time by articles published in a given year (Burrell, 2002; Glänzel, 2004). The prospective approach looks at the future, i.e., it focuses on tracking the citing papers of each cohort. We use the prospective approach as it is identified as the most straightforward approach to study these types of problems (Glänzel, 2004). Also, there is a mathematical equivalence between both approaches (Yin & Wang, 2017), so this decision does not have any impact on the final results.

We choose a cohort of scientific articles to use the prospective approach, let us say 1980. Then, we build two time series, one for the number of citations obtained in each time window (usually, six months yet, using a different time window has been shown not to change the results (Higham et al., 2017b, 2017a)) and another for the accumulated citations obtained up to a given time. In some cases, we also need to consider that external factors unrelated to obsolescence could affect our observability. For instance, in scientific papers, an artificial growth in the number of publications in Physical Review B in any year after 1980 will impact citation rates; thus, hiding the temporal decay patterns of forgetting (Higham et al., 2017b, 2017a). Therefore, to account for those external factors, we adjust the time series by normalizing them by the number of papers published in a journal each year using an inflation factor (Higham et al., 2017b, 2017a) by fixing a base year and re-scaling all the citations obtained to this base year:

$$I_f = \frac{N_j(T_0, T + \Delta T)}{N_j(T, T + \Delta T)},$$

where $N_j(T, T + \Delta T)$ is the number of papers published by the citing journal "J" at the interval $(T, T + \Delta T)$, and $N_j(T_0, T + \Delta T)$ is the number of papers published by the citing journal "J" at the interval when the paper was published. The inflation factor allows us to control external shocks and the exponential growth of science (Hall et al., 2001; Higham et al., 2017b, 2017a).

Now, we need to isolate the driving forces of     collective memory dynamics. Formally, collective attention A(t) (Fig. 2B) can be modeled as a separable function of two components (Candia et al., 2019; Candia & Uzzi, 2021; Csárdi et al., 2007; Higham et al., 2017b, 2017a; Valverde et al., 2007), and takes the form: A(t)=c(k) x S(t), where c(k) captures the effects of cumulative advantage (Fig. 2C) and S(t) captures the temporal decay (Fig. 2D).

Figure 2C shows the preferential attachment component by presenting the number of new citations ($\Delta c$) received by a paper as a function of its cumulative citations ($c(k)$), where each curve (t = 2, 3, and 5) correspond to different ages of the articles published in PRB in 1980.

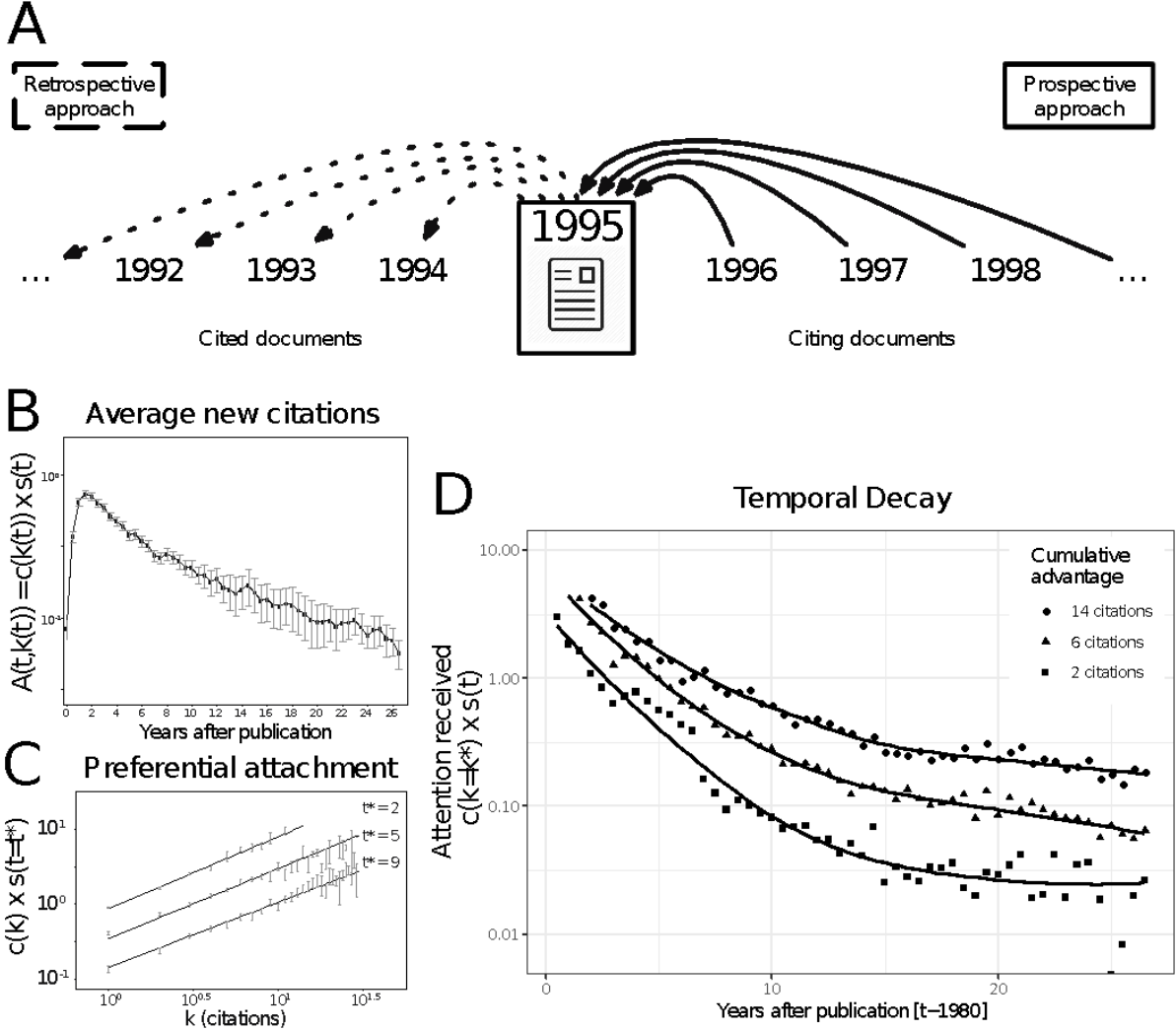

Figure 2: **A)** Scheme for the retrospective and prospective approach for tracking citations. **B)** An example of a typical adoption curve characterized by the average number of citations received each time by papers published in *Physical Review B* ($A(t)$). **C)**, Average number of citations received by a paper as a function of the cumulative citations received by that paper ($\propto c(k)$). Different curves represent different ages. **D)**, The average number of citations received by papers with the same number of cumulative citations as a function of their age ($\propto S(t)$). Different curves represent groups of papers with a different number of total citations. Note. Panels A and D are reprinted from Quantifying the Selective Forgetting and Integration of Ideas in Science and Technology by Candia & Uzzi (2021). Panels B and C are reprinted from "The universal decay of collective memory and attention", by (Candia et al., 2019).

Finally, after conditioning the time series of new citations using the accumulated citations time series figure 2D shows the temporal decay component, $S(t)$, representing the number of new citations received by papers with the same number of cumulative citations $k = k^*$ as a function of their age; that is, the dashed lines show papers for which the effect of preferential attachment is kept constant: $A(t)|_{k=k^*} \cdot c(k^*) \times S(t)$. We observe the initially fast decay followed by a slow and long-lived decline. We assume that collective attention, $S(t)$, of a cultural product is the sum of the attention it garners from both communicative acts "u" and accessing cultural records "v." Hence, at any given time $S(t) = u(t) + v(t)$. Our model, fully grounded on collective memory literature, derives a bi     exponential function for the temporal decay:

$$S(t) = N\left[e^{-(p+r)} + \frac{r}{p+r-q}\left(e^{-qt} - e^{-(p+r)t}\right)\right],$$

where S(t) quantifies the attention received —proxied as the average of new citations— conditioned to a fixed level of cumulative advantage (Fig. 2D). The parameter "p" (Fig.1) represents the decay rate of attention from communicative acts directly to oblivion. The parameter "r" (Fig. 1) illustrates the transition rate of the attention from communicative acts to cultural memory, i.e., accessing records. The parameter "q" (Fig. 1) represents the decay rate of attention from cultural memory to oblivion (for more detail about the mathematical derivation of the model, see the methods section of Candia et al., 2019).

We acknowledge that several times it is not possible to access time-series data, hindering the study of the temporal decay of collective memory and attention. Nevertheless, we can overcome this issue by introducing an inclusion criterion for cross-section data. The inclusion criterion of each cultural piece (or data point) should depend only on a proxy of its accomplishment and is, in principle, independent of the measured proxy of collective attention. For example, we cannot include biographies based on their number of language editions or their number of pageviews because our sample would be biased against old, unpopular biographies. Instead, we should include biographies of people who have accomplished similar things in a similar time, such as Olympic medalists who won a similar number of medals in the same year or basketball players who were ranked in the top rankings in the same year (Candia et al., 2019).

Similarly, to study collective attention of music, books, or movies, we should use as inclusion criteria a proxy that depends on initial accomplishments such as Billboard hot 100 ranking for songs, best sellers ranking for books, more than 1.000 votes in IMDB for movies, along with others. Then, as a general rule, the analyzed cultural pieces should be selected using a criterion that does not correlate with the collective attention measure. Thus, the inclusion criterion accounts for the cumulative advantage effect and unveils the time decay properties of collective memory.

Finally, the biexponential model predicts a transition and characteristic time when cultural memory overcomes communicative memory. In other words, the transition time quantifies for how long the communicative acts are the primary support of collective memory:

$$t_c = \frac{1}{p+r-q} log\left(\frac{(p+r)(p-q)}{r}\right)$$

Note that the transition time depends only on the model parameters (for more details about the derivation of the transition time, see the methods of     Candia et al., 2019    ).

## Concluding Remarks

Collective memory is the shared past of a group of people that shape its identity (Assmann, 2008, 2011; Candia et al., 2019; Candia & Uzzi, 2021; García-Gavilanes et al., 2017; Goldhammer et al., 1998; Halbwachs, 1997; Roediger et al., 2009; Roediger & DeSoto, 2014; Rubin, 2014; Wertsch, 2002; Zaromb et al., 2014). Communities experience two phases of memory, characterized by different levels of attention: an initial phase of a high level of attention, followed by a more extended and slower phase of forgetting. Therefore, there are three fundamental forces behind the collective memory and attention dynamics: Cumulative Advantage (attention begets attention), Communicative Acts (associated to communicative memory, i.e., oral transmission of information), and Accessing Records (related to cultural memory, i.e., the physical recording of information) (Candia et al., 2019; Candia & Uzzi, 2021; Coman, 2019). Moreover, collective memory can provide insights for building mechanistic models of the decay of collective attention.

A recent model, fully grounded on collective memory mechanisms, derives a universal function that describes the whole temporal dimension as a biexponential function. The model uncovers the communicative and cultural memory nature of the temporal dimension of collective attention dynamic, and it is empirically validated in five different cultural domains (Candia et al., 2019; Candia & Uzzi, 2021).

The biexponential model predicts a transition time between communicative and cultural memory. Given its practical applications, future research should focus on which factors could shrink or expand the transition time. For instance, extending the transition time could provide policymakers with valuable time to generate solutions and public commitment to society-wide relevant issues (Coman, 2019).

The model's parameters capture many processes; perhaps the most straightforward one is the growth in creating new cultural products that leads to a shrinking in the collective attention focused on specific cultural icons because the attention capacity of systems/communities is limited. Therefore, obsolescence makes old cultural products less engaging to people in terms of attention. We acknowledge that our aggregate approach for quantifying collective memories cannot distinguish between different forms of memory or attention loss, such as interference, suppression, or inhibition. It only provides an aggregate picture of the attention lost through these channels. Future work should focus, among others, on extending this model to an individual level, including more variables related to initial accomplishment to explain future collective attention.

## Acknowledgments

The author acknowledges the financial support of FONDECYT Project 11200986 and Data Science Institute (IDS) at Universidad del Desarrollo. Also, authors acknowledge the helpful and thorough insights of Carlos Rodriguez-Sickert and Camila Utreras.

## Further readings

- Assmann, J. (2008). Communicative and Cultural Memory. *Cultural Memory Studies. An International and Interdisciplinary Handbook*, 109–118. https://doi.org/10.1515/9783110207262.2.109

- Hirst, W., & Manier, D. (2008). Towards a psychology of collective memory. *Memory*. https://doi.org/10.1080/09658210701811912

- West, R., Leskovec, J., & Potts, C. (2021). Postmortem memory of public figures in news and social media. *Proceedings of the National Academy of Sciences*, *118*(38), e2106152118. https://doi.org/10.1073/pnas.2106152118